\documentclass[a4paper,12pt]{article}
\usepackage[stable]{footmisc}
\usepackage{amsmath,graphicx}     

\usepackage{hyperref}

\begin{document}

\begin{flushright}
  TIF-UNIMI-2024-12
\end{flushright}

\begin{center}{\Large Asymptotic Freedom in Parton Language:\\ the Birth
    of Perturbative QCD}\\
\bigskip

\vspace{0.4 truecm}

{\bf Stefano Forte}
 \\[5mm]

{\it Tif Lab, Dipartimento di Fisica, Universit\`a di Milano and\\ INFN, Sezione di Milano,\\ Via Celoria 16, I-20133 Milano, Italy\\[2mm]}

\vspace{0.8cm}

{\bf Abstract }
\end{center}
I review the contributions of Giorgio Parisi to perturbative
QCD. Concentrated in a decade, they mark the transition of the theory
of strong interactions from a set of loosely connected ideas  based on
models, to
a quantum field theory that is now an integral part of the standard model
of fundamental interactions. Parisi's contributions have established
at a very early stage ideas, methods and tools that are now standard,
and in several cases anticipated results that only became prominent in
the XXIst century.

\begin{center}
Contribution to the volume\\ {\it From Quantum Fields to Spin
  Glasses:\\ A journey through the  contributions of Giorgio Parisi to theoretical Physics}
\end{center}

\section{Parisi and QCD}
\label{sec:intro}
\bigskip

The contributions of Giorgio Parisi to the theory of strong
interactions in the perturbative domain are concentrated in one
decade: the first paper~\cite{Cabibbo:1970as} was published in 1970, and the
last~\cite{Martinelli:1979rm} in 1980. This is the decade of the
``triumph of quantum field theory'', in Sidney Coleman's
words~\cite{Coleman:1985rnk}, filled with ``wonderful things brought
back from far places to make the spectator gasp with awe and laugh
with joy''. Many of these things, in particular  those that
have to do with QCD, are  of Giorgio's own making.  

Parisi's contributions are contained in about 40 papers, out of about
a hundred he authored during this decade. His other interests included
various more formal aspects of quantum field theory, including its
first applications to statistical mechanics; the theory and
phenomenology of quantum electrodynamics, of great relevance at
Frascati, where he worked and where AdA, the first ever $e^+e^-$
collider had been conceived and built, and the ADONE collider started operating in 1969; various
aspects of hadron phenomenology and flavor physics~\cite{maianicont}; and a constant
interest in non-perturbative methods which eventually, towards the end
of the decade, developed into an interest in lattice gauge theories
that would lead to seminal contributions  a few years later~\cite{lombcont,martcont}.

These contributions  can be broadly subdivided into three partly overlapping epochs.
A first epoch, from 1970 until
about 1973, during which the main issue was understanding the meaning
of the parton model, explaining its phenomenological success, and
understanding its possible theoretical underpinnings and its
implications. In other words, understanding Bjorken scaling: the
surprising discovery that structure functions, the form factors that
describe the structure of the proton as probed by the  deep-inelastic scattering
of electrons (DIS) are
approximately scale independent. A second
epoch, ranging from 1972 to about 1976, in which the nature of QCD as a quantum field theory starts
emerging, and the parton model is understood as a limit of perturbative
QCD --- culminating with the seminal paper~\cite{Altarelli:1977zs}
that gives its name to the present paper.
In other words, understanding scaling violations: the
weak logarithmic scale dependence of hadronic structure. And a
last epoch, from 1976 to the end of the decade, in which the foundations
of perturbative QCD are laid, by introducing some of its basic ideas
and developing its fundamental tools.

The vast majority of these contributions are contained in very short
papers, with Parisi as a single author: typically two or three pages, in which a seminal idea is
presented, in the guise of a flash of light in the dark. 
The purpose of this brief contribution is to provide an overview of
this work, based on the full list of publications of Giorgio Parisi,
by discussing most, if not all of the relevant contributions, in an
attempt of capturing some of this light.

\section{Scaling and the parton model}
\bigskip
\label{sec:parton}

The earliest Parisi paper on the phenomenology of the strong interactions~\cite{Cabibbo:1970mh}  (and
his third published paper on record ---
the two previous one dealing with quantum field theory topics)
concerns what in many current standard textbooks is indeed the first
application of QCD to be discussed (see
e.g. Refs.~\cite{Peskin:1995ev,Ellis:1996mzs}): the computation of the
total rate of production of hadrons at an electron positron-collider,
now usually presented as the $R$ ratio
\begin{equation}
  R=\frac{\sigma(e^+e^-\to\hbox{hadrons})}{\sigma(e^+e^-\to\mu^+\mu^-)}
, \end{equation} 
i.e.\ the ratio of hadron to muon-antimuon production cross-sections. 
Equation~(1) of
this paper states the now well-known leading-order result: $R=\sum_i Q^2_i$,
where $Q_i$ are the electric charges of partons, i.e.\ the constituents
of the proton as probed in the perturbative domain, and the sum runs
over all constituents, and thus can be used to establish both the
nature and the
total number of constituents --- and thus eventually provided evidence
for the existence of three quark colors.

The derivation is quite close to the modern textbook
derivation~\cite{Peskin:1995ev}, which is  based on
the Wilson expansion: indeed, the result is obtained by writing the
cross-section in terms of a current matrix element, and evaluating the
latter as a vacuum expectation value. The dominance of the 
contribution from operators with the lowest dimension,
that in modern language follows from the Wilson
expansion, here instead is justified using an argument that captures the basic
underlying physics, namely, the scale hierarchy between  the hard
$e^+e^-$ annihilation, and the softer dynamics of the hadronization.
The argument was previously proposed in the context of
DIS~\cite{Drell:1969ca}
and is thereby generalized to
$e^+e^-$ annihilation.

The physics question that underlies this paper is made explicit in the
title of a companion paper~\cite{Cabibbo:1970as}: what is the nature
of partons? Are they quarks and other ``mythical'' (sic) constituents
of hadrons? Or are they just hadrons? In the latter case, it is
pointed out that DIS would be probing the structure of the vacuum, not
of hadrons: the virtual photon would be creating a hadron-antihadron
pair out of the vacuum. The paper explores this latter option, and it
shows how it could be tested experimentally, by measuring the ratio of
longitudinal to transverse cross-sections, which in this case could be determined
using SU(3) considerations. 

The tension between the parton model viewpoint and the opposite
option of computing in terms of hadronic states is bypassed by taking a
``duality'' approach  in Ref.~\cite{Parisi:1972rh}, i.e.\ assuming that
results can be equivalently obtained in terms of partons or
resonances.  Combining this assumption with SU(3) flavor symmetry
considerations leads to the sum rule
\begin{equation}\label{eq:gottfried}
  \int_0^1\! \frac{dx}{x} \left[F_2^p(x)-F_2^n(x)\right] =\frac{1}{3}.
  \end{equation}
for the proton-neutron difference of the $F_2$ structure
functions that parametrize the deep-inelastic
cross-section. Amusingly, in more recent textbooks (see 
e.g. Ref.~\cite{Roberts:1990wwe}) this result is  derived using naive parton
model arguments, and referred  to as the Gottfried~\cite{Gottfried:1967kk}
sum rule --- though it is nowhere to be found in Gottfried's
paper~\cite{Gottfried:1967kk}. The discovery of its experimental violation in the early
nineties arguably opened up the modern era of understanding of the parton
substructure of the nucleon~\cite{Martin:1990vy}.

The ultimate underlying question is of course whether Bjorken scaling
and the parton model could be justified theoretically. The question is
addressed directly in Ref.~\cite{Parisi:1972kji}, in which Bjorken
scaling is derived using renormalization group arguments,
from the Wilson expansion and the Callan-Symanzik
equation, under the assumption that the $\beta$ function of the strong
interaction Lagrangian has a nontrivial
fixed point, where the theory becomes conformal. The derivation is
essentially identical to the standard 
current textbook argument, except that the latter is based on
expanding perturbatively in the coupling about zero, so scaling only
holds asymptotically and the parton model receives logarithmic
corrections --- as Parisi's work would fully elucidate only a few years later.

Before turning to the line of thought that led to this celebrated
result, it is worth spending a few words to discuss some studies 
that were part of investigations of QED, rather than QCD, but are
precursors of techniques that are now part of the standard
QCD toolbox, and indeed some of which played an important role in
subsequent QCD studies of Giorgio's. The first~\cite{Parisi:1971qe} is a discussion of the $e^+e^-\to
e^+e^-\gamma$ process in QED in the hard Bremsstrahlung limit, in
which the photon energy is large. This paper is based on the use of
the
Weizs\"acker-Williams, or
quasi-real photon
approximation, in which the photon emission rate is computed in
terms of a universal probability for the electron to split into a
photon. This approximation will later play an important role
in the derivation of evolution equations for parton distributions,
namely, 
the Altarelli-Parisi equations, that I will discuss in
Sect.~\ref{sec:ap} below. In this paper, the result found using this
approximation is compared
to the exact result. The approximation 
holds in the limit of collinear photon emission, but
it is shown to provide an 
acceptable approximation for surprisingly large angles. With
hindsight, this explains the unexpected phenomenological success
of the Altarelli-Parisi equation, and consequently of Bjorken scaling,
even at
the relative low scales that were experimentally probed at that time,
a phenomenon that was later called precocious scaling.

The second is the companion paper~\cite{Parisi:1971tb}, in which the
exact computation for this QED process is presented. In this work it is observed that, when
computing Feynman diagrams with several particles in the final state,
it is highly advantageous to use, instead of Dirac traces,  
the helicity amplitude method, previously introduced in Ref.~\cite{Parisi:1972cp}. The  
computation is then performed using this  method, now standard,  but only
popularized by others many years later~\cite{Mangano:1990by}. The
computational technique itself is in fact the main point
of the paper, whose stated goal in its abstract is to ``propose a general method to evaluate Feynman
graphs'' which ``avoids the calculation of very long traces''.

\section{Scaling violations}
\label{sec:ap}
\bigskip

Only three months (based on submission dates) elapsed between the
proof that Bjorken scaling holds in the presence of a nontrivial fixed
point~\cite{Parisi:1972kji}, and the first exploration of perturbative
scaling violations~\cite{Parisi:1973nx}. Indeed, it had been meanwhile
realized that Bjorken scaling is only approximate: deep-inelastic
observables are not scale independent, rather, they display a
weak logarithmic dependence on the scale $Q^2$ at which the target
proton is probed, rather that the strong power-like
behavior that one might expect based on naive dimensional
analysis. The name of the game thus had changed from understanding
scaling, to understanding the nature of these logarithmic scaling violations.

The paper~\cite{Parisi:1973nx}, with the
seemingly modest goal of ``showing how to use the data on
deep-inelastic scattering to put bounds on the values of anomalous
dimensions of the operators involved in the Wilson expansion of the
product of two currents near the light-cone'', makes a striking
observation. Namely, that if a moment (integral)
of a structure functions displays a scaling law
\begin{equation}\label{eq:scal}
  \int_0^1\! dx\, x^{N-1}F_2(x,Q^2)= 
  \left(\frac{Q^2}{\Lambda^2}\right)^{-\alpha\gamma_N},
\end{equation}
characterized by an anomalous dimension $\alpha\gamma_N$,
(where I have used modern notation), then the convolution theorem for
Mellin transforms implies that 
\begin{equation}\label{eq:mellinv}
F_2(x,Q^2)=F_2(x,Q_0^2)-\alpha \int_x^1\frac{dy}{y}
 P\left(\frac{x}{y}\right)\ln\frac{Q^2}{Q_0^2}  F_2(y,Q_0^2)+ O(\alpha^2),
\end{equation}
where the anomalous dimension is the Mellin transform of  $P(x)$:
\begin{equation}\label{eq:spldef}
\gamma_N = \int_0^1 dx\, x^{N-1} P(x).
\end{equation}
In other words, the finite scale dependence Eq.~(\ref{eq:scal}), in
which $\Lambda$ plays the role of an initial condition, emerges as
the consequence of an infinitesimal scale transformation from $Q_0$ to
$Q$, given by Eq.~(\ref{eq:mellinv}).

This result is used to get an estimate of the size of the
coefficient $\alpha$ (which is viewed as a normalization, rather than
a coupling), based on an assumed plausible form of the dependence of
the
anomalous
dimension $\gamma_N$ on $N$. The truly astonishing
nature of the paper is the fact that Eq.~(\ref{eq:mellinv}) appears
as the consequence of the  somewhat casual side-remark, that the
convolution theorem for Mellin transforms implies that
a scaling law of the form of 
Eq.~(\ref{eq:scal}), which is obeyed by moment integrals of the
structure function $ F_2(x,Q^2)$, can be cast in equivalent
form as a scaling law satisfied by the structure function
itself, governed by a kernel implicitly defined by
Eq.~(\ref{eq:spldef}). At this time,  this might appear as a purely mathematical
trick, but it effectively contains the basic physics of
the Altarelli-Parisi equation, that we will soon get to. Namely, that
the kernel $P(x)$ --- the Altarelli-Parisi 
splitting function, in modern terminology --- provides ``the violation in
each point of Bjorken scaling law'', i.e.
the local scaling violation for
each value of $x$, rather than that of the integral of the structure
function over $x$. 

The question of how to actually compute the anomalous dimension itself
is addressed in Ref.~\cite{Parisi:1973xn}, in which it is shown that
the scaling violations of the moments of structure functions that
appear on the
left-hand side of Eq.~(\ref{eq:scal}) are governed
by the anomalous dimensions of the matrix elements of operators with
the lowest dimension and increasing spin, i.e.\ leading-twist operators (again using
modern terminology). These  are in turn computed in $\phi^4$
theory: the {\it tour de force }calculation is based on solving the Dyson equation
satisfied by the operator vertex, rather than using the Wilson
expansion as in the modern textbook approach.

A relevant observation is that all these developments predate QCD as a
quantum field theory of the strong interaction, and in particular they predate
the discovery of the $J/\psi$ in 1974, the so-called November
revolution~\cite{Gross:2022hyw} that convinced the community that QCD
is the correct theory of the strong interactions. Their significance became
clear once a non-abelian gauge theory  was accepted as the quantum field
theory of the strong interaction, and it was realized that this theory
is amenable to a
perturbative treatment in the large-momentum limit.

Indeed, in Ref.~\cite{Parisi:1974sq} the mathematical insight of
Ref.~\cite{Parisi:1973nx} and  Eq.~(\ref{eq:mellinv}) is combined with
the newly acquired knowledge from the seminal work of Gross and
Wilczeck~\cite{Gross:1974cs} on anomalous dimensions and scaling laws
of moments of structure functions, to show that structure
functions satisfy an integro-differential evolution equation that
determines their scale dependence. This equation takes the form (in modern notation)
\begin{equation}\label{eq:dglap0}
\frac{\partial}{\partial \ln Q^2} F^{p-n}_2(x,Q^2)=\alpha_s(Q^2)\int_x^1 dy\,
  P\left(\frac{x}{y}\right)  F_2^{p-n}(y,Q^2),
\end{equation}
where  $F^{p-n}_2$ is the difference between proton and neutron
deep-inelastic structure functions, $\alpha_s(Q^2)\propto\frac{1}{\ln Q^2}$ is an asymptotically vanishing coupling, and $P(z)$ is related
through Eq.~(\ref{eq:spldef}) to the nonsinglet anomalous dimension
that had been computed by Gross and Wilczek. Amusingly, this paper still predates by
several months the discovery of the  $J/\psi$.
 Once again, as in
Ref.~\cite{Parisi:1973nx}, the fact that the scaling law satisfied by
moments of structure functions can be equivalently viewed as the
consequence of an evolution equation satisfied by the structure
function itself, with an evolution kernel related by Mellin
transformation to the scaling anomalous dimension, is presented as a
mathematical observation. In the words of Ref.~\cite{Parisi:1974sq}, the scaling law and the evolution equation
``are mathematically equivalent, however we believe that the second
equation is easier to test''. In fact, in this work and subsequent
investigations based on it~\cite{Altarelli:1976ms,Parisi:1976fz} the
emphasis is on exploiting these results to obtain testable predictions.

However, at this point the step to realizing that
this mathematical observation actually has a profound physical meaning
--- the step leading to the
Altarelli-Parisi equation --- is
very short indeed. A preview is given in the 1976 Moriond lectures,
Ref.~\cite{Parisi:1976qj}, whose abstract states ``the theory of
scaling violations in deep inelastic  scattering is presented using
the parton model language''. In these lectures, evolution equations of
the form of Eq.~(\ref{eq:dglap0})
are derived in quantum electrodynamics  (QED) for the ``constituents
of the electron''. Namely, it is observed that an electron can radiate
photons that in turn radiate electron-positron pairs, so when probing
an electron one is actually probing its internal structure in terms of
photons, electrons and positrons. By observing that (at leading
perturbative order) the structure function   $F^{p-n}_2(x,Q^2)$ in
Eq.~(\ref{eq:dglap0}) is proportional to the number density of
electrically charged
constituents   of the target electron,
the evolution equation
Eq.~(\ref{eq:dglap0}) is then interpreted as a consequence of the fact
that this number density, i.e.\ the internal structure of the
electron, depends on the scale at which the electron is probed
because of the scale dependence induced by this radiation process.

The computation is performed  using the Weizs\"acker-Williams (or
equivalent photon) approximation that was
already encountered in Ref.~\cite{Parisi:1971qe}. The
Weizs\"acker-Williams method shows that quantum interference in the
radiation process is
suppressed by inverse powers of the radiation scale, so, for QED, it
is suppressed by powers of
$\frac{m^2_e}{Q^2}$, where $m_e$ is the electron mass, and  $Q^2$ is the scale at which the electron
is probed. Consequently the interaction rate --- the photon-electron
cross-section in this case --- factorizes in terms of an elementary
photon-electron cross-section,
times a universal radiation rate that does not depend on the specific
process but rather is a property of the target (the electron, in this
case) and can be expressed as a radiation pseudo-probability --- not
quite a probability because in general it is not positive definite.

It is then observed that if
one tries to apply the same reasoning to QCD, namely to the structure
of the proton, rather than the electron, there is a fundamental
difference, namely, that in QED the vacuum behaves as a dielectric,
and consequently the electric charge is screened at large distances,
and   increases at shorter distances (an argument that
is now textbook~\cite{Peskin:1995ev}). In QCD the opposite
happens:  strongly interacting matter is, in the words of
Ref.~\cite{Parisi:1976qj},  an ``enantion'',
loosely translatable as a ``contrarian'': it anti-screens, and consequently the charge
decreases at shorter distances --- asymptotic freedom. Combining these
two physical pictures, scale dependence from radiation, and asymptotic
freedom from anti-screening, a simple physical picture of the scaling
laws of proton structure in QCD  emerges.
Hence, the key insight in
these lectures is to interpret the scaling laws obeyed by
deep-inelastic structure function as a consequence of the scaling law
of the number densities of partons --- the proton's constituents. This then provides a clear
physical picture of scale dependence from evolution equations
satisfied by structure functions  as a showering process, as
realized in modern Monte Carlo parton shower
codes~\cite{Campbell:2022qmc}.
The solution to these evolution equations corresponds to summing a
class of Feynman diagrams that had been considered previously by Gribov and Lipatov,
who had studied  deep-inelastic scattering in a
theory with massive vector mesons
a few years earlier~\cite{Gribov:1972ri}. However, the derivation
presented here is completely independent, and in fact the physical
picture of evolution equations is missing in the Gribov-Lipatov
approach.

All the pieces of the puzzle come together in the
subsequent celebrated, seminal
Altarelli-Parisi paper~\cite{Altarelli:1977zs} (of which the Moriond
lectures~\cite{Parisi:1976qj} are presented as a ``preliminary, less
complete version''). This is of course one
of the most widely cited papers in high-energy physics. In this paper
the focus shifts from the deep-inelastic structure function, to the
structure of the proton itself. Indeed, the first insight  is that the scale dependence of structure function is a
consequence of an underlying scale dependence of the structure of the
proton, as reflected by parton distribution functions, that provide
number densities for the constituents of the proton. The second
insight is that the previous mathematical observation~\cite{Parisi:1973nx,Parisi:1974sq}, that
scaling laws  can be obtained by solving an integro-differential
equations and taking the Mellin transform of the result, can
be combined with the physical idea that the proton constituents can be
viewed as composite systems, an idea previously
suggested by Kogut and Susskind~\cite{Kogut:1974ni}.

The combination of these two insights is made quantitative by
explicitly computing the probability density for finding a constituent
of the proton inside another constituent, i.e.\ the probability for
a parton to split into two. The latter is simply given by the
universal rate of emission of an extra final-state parton from an
initial state one, which can be computed, as  I have already discussed,
using the Weizs\"acker-Williams
approximation, in terms of ``splitting functions'' that are completely
determined by transition matrix elements from the initial parton to
the final parton pair. The relevant emission matrix elements are determined
using the so-called  
``old'', i.e.\ time-ordered perturbation theory, a method that is
rarely~\cite{Sterman:1993hfp} discussed in modern textbook treatments, and
that has been recently revived~\cite{Borinsky:2022msp}. 

The final result is a pair of coupled evolution equations for parton
distributions of quarks of the $i$-th flavor or antiflavor
$q^i(x,t)$ and gluons $g(x,t)$, with $t=\ln(Q^2/Q_0^2)$, $x$
the fraction of the parent proton energy-momentum carried by the given
parton, $Q^2$  the scale at which the proton is probed, and $Q_0^2$
an arbitrarily chosen reference scale. These have a form that is
entirely analogous to that of Eq.~(\ref{eq:dglap0}), namely
\begin{align}\label{eq:dglap}
  \frac{d}{dt} q^i(x,t)
  &=\frac{\alpha_s(t)}{2\pi}\int_x^1 \frac{dy}{y}
    \left[ P_{qq}\left(\frac{x}{y}\right)q^i(y,t)+ P_{qg}\left(\frac{x}{y}\right)g(y,t) \right],\\
  \frac{d}{dt} g(x,t)
  &=\frac{\alpha_s(t)}{2\pi}\int_x^1 \frac{dy}{y}
  \left[ P_{gq}\left(\frac{x}{y}\right)q^i(y,t)+
    P_{gg}\left(\frac{x}{y}\right)g(y,t)\right],
\end{align}
where the splitting functions $P_{ij}(z)$ express the probability to find
a parton (quark or gluon) $i$ inside a parton $j$ with fraction $z$ of
the parent parton's momentum.
These are the celebrated Altarelli-Parisi equations. The same
equations are then derived also in the case of polarized parton
distributions, which express the probability of the parton to carry
the parent proton's spin. 

The derivation of
these equations presented in Ref.~\cite{Altarelli:1977zs} is arguably
rather more transparent than the majority of current available
textbook treatments. However, their significance goes well beyond the
technical aspects of their derivation. In fact, the fundamental meaning of
this paper is sometimes not fully appreciated.
Indeed, this work is sometimes
described~\cite{Peskin:1995ev} as an independent re-derivation of the
previous argument by
Gribov and Lipatov~\cite{Gribov:1972ri}, that shows how to obtain
evolution equations for deep-inelastic structure functions themselves,
rather than their moments.
Whereas of course the Altarelli-Parisi paper does
contain, as it is stated in its introduction ``an alternative
derivation of all results of current interest for the $Q^2$ behavior
of deep inelastic structure functions'' (alternative to the derivation
based on the scale dependence of matrix elements of leading-twist
operators, that is), this is
not the main point. Rather, the point, as stated
in the title of the paper, is to express quantum-field theoretical results of QCD in
parton language. In fact, this work brings to completion
the research program and line of thought that started in the early
papers that asked the questions: what is the nature of partons? and
why does the parton model work?

Indeed, even in modern textbook treatments~\cite{Ellis:1996mzs} a
parton model language and parton-based approach 
to QCD factorization is often taken for granted: it is assumed that hard
processes can be described by starting with the parton model, and
computing perturbative corrections to it. Yet, the theoretical
justification for this is unclear --- it is unclear why the simple
assumptions of the parton model are a reasonable starting
point\footnote{See e.g.\ Sect.~9.11 of Ref.~\cite{Collins:2011zzd} for
a modern critique of a naive parton-based approach.}. What the
Altarelli-Parisi paper shows, is that, for deep-inelastic structure
functions, a parton-based computation is mathematically equivalent to
the result found starting from the Wilson expansion, and then solving
renormalization-group equations for Wilson coefficients. Indeed, a
terse comment in the conclusion explains ``the reasons for the success
of this simple method'' --- the renormalization group equations. So
what the Altarelli-Parisi paper amounts to, is a rigorous proof of
QCD factorization, albeit in the limited case of deep-inelastic
scattering. The paper is at the basis of the modern description of
the structure of the proton in terms of parton distributions (see
e.g. Ref.~\cite{NNPDF:2021njg} and ref.\ therein) but its true significance
is in establishing perturbative factorization.

\section{The birth of QCD phenomenology}
\label{sec:pheno}
\bigskip

With QCD factorization on a solid field-theoretic basis, at least for
deep-inelastic scattering, the leap to 
full-fledged QCD phenomenology was the logical development. This was
laid out in a series of papers, submitted within the
short span of  two years, which set the basis of QCD
calculations for hadronic
processes~\cite{Altarelli:1977kt,Altarelli:1978pn}, the
definition of jet substructure observables~\cite{Parisi:1978eg,Martinelli:1979rm},
the determination of parton
distributions~\cite{Parisi:1978jv}, the scale dependence of fragmentation functions~\cite{Parisi:1978kw},   
transverse momentum resummation~\cite{Parisi:1979se}, and threshold
resummation~\cite{Parisi:1979xd}.
Many of these contributions anticipate ideas and methods that have
marked the development of QCD phenomenology, sometimes several decades
later.

Some of these papers had a large impact, and shaped subsequent
developments, though some were only recognized at a later stage, 
and some not at all. In particular, Ref.~\cite{Altarelli:1977kt}
presents a first computation of the dimuon transverse momentum distribution
in Drell-Yan production. The celebrated
Collins-Soper-Sterman (CSS) work~\cite{Collins:1984kg}, that later
set the modern
standard for the perturbative calculation for this process,
refers to Ref.~\cite{Altarelli:1977kt} as the foundation of
the whole field: ``with the advent of QCD
factorization one was able to write the DY cross-section at measured
$Q_T$''. Interestingly, the subsequent paper
Ref.~\cite{Altarelli:1978pn} considers the effect of taking into
account the 
transverse momentum of partons, a topic that
has only become fashionable in this century, with studies of
transverse-momentum dependent parton distributions (TMDs~\cite{Angeles-Martinez:2015sea}).

The leading-order computation of Ref.~\cite{Altarelli:1977kt}
immediately raised the question of the instability of its result
towards the inclusion of
logarithmically enhanced perturbative corrections in the small $p_T$
region. The all-order resummation of these terms is accomplished in
Ref.~\cite{Parisi:1979se}, by observing that they stem
from the emission of soft
(i.e.\ very low energy) gluons, whose contribution was
known~\cite{Altarelli:1979ub} to exponentiate.  Again, in  
CSS~\cite{Collins:1984kg} this later paper is  described with the
words ``These authors introduced more powerful techniques: they
worked with the Fourier transform with respect to $Q_T$\dots and they
showed the usefulness of soft gluon methods.'' 
The same soft gluon methods, and in particular the powerful idea
of performing resummation in terms of a variable which is conjugate
by integral transform (Mellin transform, in this case) to the physical
variable whose soft limit is being considered, are used in
Ref.~\cite{Parisi:1979xd} to perform leading-log threshold resummation of the DIS
and DY cross-sections, a result only extended to next-to-leading log
about ten years later~\cite{Sterman:1986aj,Catani:1989ne}.

In fact, in the brief space of two pages the single-author
Ref.~\cite{Parisi:1979xd} manages to contain the following insights,
for which essentially no proof is given:
the kinematic origin of the different behavior of the DY and DIS cases
(proven in detail in Ref.~\cite{Forte:2002ni}), the impact of
running-coupling effects on resummation (fully proven in
Ref.~\cite{Contopanagos:1996nh}), and finally the impact of the analytic
continuation from spacelike to timelike in relating DY to DIS, fully
studied more than ten years later in Ref.~\cite{Magnea:1990zb}, and whose phenomenological consequences were re-discovered 
in Ref.~\cite{Ahrens:2008qu}.

Separate lines of developments are related to the needs of precision
phenomenology, thereby anticipating the LHC era. In
Ref.~\cite{Parisi:1978jv} it is suggested that an accurate
parametrization of parton distributions requires extending the
standard parametrization that reproduces the limiting behaviors at
large and small momentum fractions with bases of orthogonal
polynomials, something that groups still using this kind of 
parametrization first started doing only with the advent of the
LHC~\cite{Martin:2012da}.

In Ref.~\cite{Parisi:1978eg} the energy-momentum correlation of pairs
of hadrons in the final state of a collision (hadron-hadron,
lepton-hadron, or lepton-lepton) is
introduced. This is the precursor of  event shape
variables~\cite{Ellis:1996mzs}:  this work is typically
cited as the origin of the so-called $C$-parameter, and energy
correlators in particular have recently become an extremely fashionable subject (see
e.g.\ Ref.~\cite{Gao:2023ivm} and ref.\ therein). However, it is interesting
to observe that the idea here is somewhat different: namely,
to define the observable in the whole
(``superinclusive'') final state, rather than inside a jet. This
leads to a quantity that can be computed purely using
renormalization-group arguments, a suggestion that to the best of
our knowledge has never been pursued. A subsequent paper in which 
a jet structure observable is computed perturbatively~\cite{Martinelli:1979rm}
is one of only eleven papers out of the 230 published by Parisi to
this day to have received no citations.

A similarly limited impact is that of Ref.~\cite{Parisi:1978kw}, which
presents a first computation of the scale dependence of fragmentation
functions. These  describe the rate at which partons can fragment into
final-state hadrons and are thus the timelike counterpart of the
parton distributions which describe the rate at which partons can be
pulled out of an initial-state hadron; they are also currently a
fashionable topic in view of their future measurement at the forthcoming
Electron-Ion Collider (see e.g.\ Ref.~\cite{Bonino:2024adk} and
ref.\ therein). Their leading-order scale dependence 
is also governed by the Altarelli-Parisi equation, as
this paper first showed. The paper, however,
collected only four citations, despite being almost
contemporary to the work that is usually presented as the first
derivation of this result~\cite{Owens:1978qz}.

\section{The work and its context}
\bigskip
\label{sec: conc}

The last paper by Giorgio Parisi dealing with perturbative QCD is
Ref.~\cite{Parisi:1979xd}, published nine and a half years after the
first~\cite{Cabibbo:1970mh}. Parisi's QCD  work in its complex is specifically characterized
by a striking feature, namely, the desire to understand information
coming from experiments, specifically with the language of quantum
field theory, and to turn this understanding into testable
predictions. Despite their brevity,  essentially all of the papers discussed here end with a
quantitative estimate, or a brief discussion of phenomenology.
Indeed, quantum field theory is the common thread  that connects all
papers published by Parisi during this decade,  ranging from its more
formal aspects, to its applications to particle physics and later also
statistical mechanics and condensed matter phenomenology, with
non-perturbative aspects of field theory providing the other strong
connection to phenomenology.

It is indeed the transition from perturbative to non-perturbative,
characterized by studies of the divergence of the perturbative
expansion in field theory~\cite{brezcont}, and the nonperturbative study of field
theory on the lattice~\cite{lombcont,martcont}, that gradually became the dominant themes of
research towards the end of the decade. And indeed in the subsequent
decade Parisi's work shifted mostly to lattice field theory, and also
statistical field theory~\cite{brezcont,vulpcont,rychcont}, thereby completing the transition from
perturbative to non-perturbative of his line of thought.

Within the more general context of research in perturbative QCD,
Parisi's contribution was instrumental in establishing quantum field
theory as the paradigm of the theory of fundamental interactions, and
quantum chromodynamics as the paradigm of quantum field theory.

\section*{Acknowledgments and disclaimer}
I am grateful to the organizers of the seminar series ``The 
interdisciplinary contribution of Giorgio Parisi to theoretical
physics'', that took place at Sapienza University of Rome in 2022-2023,
for inviting me to give a talk in the series,
and for subsequently taking the initiative
of collecting in a volume the write-ups of all talks. Unlike most of
the other contributors, I did not have the privilege of collaborating
with Giorgio directly. Somehow, despite my longtime association
with my late teacher, mentor and friend Guido Altarelli, I never managed to
discuss  his joint work with Parisi with him. Moreover, Parisi's QCD work was
done at a time when I was a school kid, so I could not witness its
development first-hand. I consequently had to approach it with the
attitude of a historian (which I am not), and to some extent of a
teacher (which I certainly am). I thus have the obligation of
apologizing in advance for any misunderstanding of which I might well
have been responsible. Among the organizers, I am especially grateful to Marco Bonvini and Maria
Chiara Angelini for comments and criticism on a preliminary version of
this manuscript. I also thank Giovanni Stagnitto for spotting several
typos.
I would finally like to thank Giorgio for
graciously sitting through my talk and sharing with me some comments,
thoughts and anecdotes.

\bibliographystyle{UTPstyle}
\bibliography{parisi}

\providecommand{\href}[2]{#2}\begingroup\raggedright\begin{thebibliography}{10}

\bibitem{Cabibbo:1970as}
N.~Cabibbo, G.~Parisi, M.~Testa, and A.~Verganelakis, {\it {Deep-inelastic
  scattering and the nature of partons}},  {\em Lett. Nuovo Cim.} {\bf 4S1}
  (1970) 569--574.

\bibitem{Martinelli:1979rm}
G.~Martinelli and G.~Parisi, {\it {Testable QCD predictions for sphericity-like
  distributions in $e^+ e^-$ annihilation}},  {\em Phys. Lett. B} {\bf 89}
  (1980) 391--393.

\bibitem{Coleman:1985rnk}
S.~Coleman, {\em {Aspects of Symmetry}: {Selected Erice Lectures}}.
\newblock Cambridge University Press, Cambridge, U.K., 1985.

\bibitem{maianicont}
L.~Maiani, {\it {Contribution to this volume}}, .

\bibitem{lombcont}
M.~P. Lombardo, {\it {Contribution to this volume}}, .

\bibitem{martcont}
G.~Martinelli, {\it {Contribution to this volume}}, .

\bibitem{Altarelli:1977zs}
G.~Altarelli and G.~Parisi, {\it {Asymptotic Freedom in Parton Language}},
  {\em Nucl. Phys. B} {\bf 126} (1977) 298--318.

\bibitem{Cabibbo:1970mh}
N.~Cabibbo, G.~Parisi, and M.~Testa, {\it {Hadron Production in e+ e-
  Collisions}},  {\em Lett. Nuovo Cim.} {\bf 4S1} (1970) 35--39.

\bibitem{Peskin:1995ev}
M.~E. Peskin and D.~V. Schroeder, {\em {An Introduction to quantum field
  theory}}.
\newblock Addison-Wesley, Reading, USA, 1995.

\bibitem{Ellis:1996mzs}
R.~K. Ellis, W.~J. Stirling, and B.~R. Webber, {\em {QCD and collider
  physics}}, vol.~8.
\newblock Cambridge University Press, 2, 2011.

\bibitem{Drell:1969ca}
S.~D. Drell, D.~J. Levy, and T.-M. Yan, {\it {A Field Theoretic Model for
  electron-Nucleon Deep-Inelastic Scattering}},  {\em Phys. Rev. Lett.} {\bf
  22} (1969) 744--748.

\bibitem{Parisi:1972rh}
G.~Parisi, {\it {A duality sum rule for deep inelastic scattering}},  {\em
  Lett. Nuovo Cim.} {\bf 3S2} (1972) 395--396.

\bibitem{Roberts:1990wwe}
R.~G. Roberts, {\em {The Structure of the proton: Deep-inelastic scattering}}.
\newblock Cambridge Monographs on Mathematical Physics. Cambridge University
  Press, 2, 1994.

\bibitem{Gottfried:1967kk}
K.~Gottfried, {\it {Sum rule for high-energy electron - proton scattering}},
  {\em Phys. Rev. Lett.} {\bf 18} (1967) 1174.

\bibitem{Martin:1990vy}
A.~D. Martin, W.~J. Stirling, and R.~G. Roberts, {\it {Parton distributions,
  the Gottfried sum rule and the W charge asymmetry}},  {\em Phys. Lett. B}
  {\bf 252} (1990) 653--656.

\bibitem{Parisi:1972kji}
G.~Parisi, {\it {Bjorken scaling and the parton model}},  {\em Phys. Lett. B}
  {\bf 42} (1972) 114--116.

\bibitem{Parisi:1971qe}
G.~Parisi and F.~Zirilli, {\it {Hard bremsstrahlung in e+ e- collisions}},
  {\em Lett. Nuovo Cim.} {\bf 2S2} (1971) 395--396.

\bibitem{Parisi:1971tb}
G.~Parisi and F.~Zirilli, {\it {Angular correlations of the decay products of
  two heavy leptons}},  {\em Lett. Nuovo Cim.} {\bf 2S2} (1971) 775--776.

\bibitem{Parisi:1972cp}
G.~Parisi and F.~Zirilli, {\it {A simple method for computing electrodynamic
  processes of high order}},  {\em Nuovo Cim. A} {\bf 11} (1972) 37--44.

\bibitem{Mangano:1990by}
M.~L. Mangano and S.~J. Parke, {\it {Multiparton amplitudes in gauge
  theories}},  {\em Phys. Rept.} {\bf 200} (1991) 301--367,
  [\href{http://xxx.lanl.gov/abs/hep-th/0509223}{{\tt hep-th/0509223}}].

\bibitem{Parisi:1973nx}
G.~Parisi, {\it {Experimental limits on the values of anomalous dimensions}},
  {\em Phys. Lett. B} {\bf 43} (1973) 207--208.

\bibitem{Parisi:1973xn}
G.~Parisi, {\it {How to measure the dimension of the parton field}},  {\em
  Nucl. Phys. B} {\bf 59} (1973) 641--646.

\bibitem{Gross:2022hyw}
F.~Gross et~al., {\it {50 Years of Quantum Chromodynamics}},  {\em Eur. Phys.
  J. C} {\bf 83} (2023) 1125, [\href{http://xxx.lanl.gov/abs/2212.11107}{{\tt
  arXiv:2212.11107}}].

\bibitem{Parisi:1974sq}
G.~Parisi, {\it {Detailed Predictions for the $p-n$ Structure Functions in
  Theories with Computable Large Momenta Behavior}},  {\em Phys. Lett. B} {\bf
  50} (1974) 367--368.

\bibitem{Gross:1974cs}
D.~J. Gross and F.~Wilczek, {\it {Asymptotically Free Gauge Theories. 2.}},
  {\em Phys. Rev. D} {\bf 9} (1974) 980--993.

\bibitem{Altarelli:1976ms}
G.~Altarelli, G.~Parisi, and R.~Petronzio, {\it {Charmed Quarks and Asypmtotic
  Freedom in Neutrino Scattering}},  {\em Phys. Lett. B} {\bf 63} (1976)
  183--187.

\bibitem{Parisi:1976fz}
G.~Parisi and R.~Petronzio, {\it {On the Breaking of Bjorken Scaling}},  {\em
  Phys. Lett. B} {\bf 62} (1976) 331--334.

\bibitem{Parisi:1976qj}
G.~Parisi, {\it {An Introduction to Scaling Violations}},  in {\em {11th
  Rencontres de Moriond}: {weak interactions and neutrino physics}},
  pp.~83--114, 4, 1976.

\bibitem{Campbell:2022qmc}
J.~M. Campbell et~al., {\it {Event generators for high-energy physics
  experiments}},  {\em SciPost Phys.} {\bf 16} (2024), no.~5 130,
  [\href{http://xxx.lanl.gov/abs/2203.11110}{{\tt arXiv:2203.11110}}].

\bibitem{Gribov:1972ri}
V.~N. Gribov and L.~N. Lipatov, {\it {Deep-inelastic $e p$ scattering in
  perturbation theory}},  {\em Sov. J. Nucl. Phys.} {\bf 15} (1972) 438--450.

\bibitem{Kogut:1974ni}
J.~B. Kogut and L.~Susskind, {\it {Scale invariant parton model}},  {\em Phys.
  Rev. D} {\bf 9} (1974) 697--705.

\bibitem{Sterman:1993hfp}
G.~F. Sterman, {\em {An Introduction to quantum field theory}}.
\newblock Cambridge University Press, 8, 1993.

\bibitem{Borinsky:2022msp}
M.~Borinsky, Z.~Capatti, E.~Laenen, and A.~Salas-Bern\'ardez, {\it
  {Flow-oriented perturbation theory}},  {\em JHEP} {\bf 01} (2023) 172,
  [\href{http://xxx.lanl.gov/abs/2210.05532}{{\tt arXiv:2210.05532}}].

\bibitem{Collins:2011zzd}
J.~Collins, {\em {Foundations of Perturbative QCD}}, vol.~32 of {\em Cambridge
  Monographs on Particle Physics, Nuclear Physics and Cosmology}.
\newblock Cambridge University Press, 7, 2023.

\bibitem{NNPDF:2021njg}
{\bf NNPDF} Collaboration, R.~D. Ball et~al., {\it {The path to proton
  structure at 1\% accuracy}},  {\em Eur. Phys. J. C} {\bf 82} (2022), no.~5
  428, [\href{http://xxx.lanl.gov/abs/2109.02653}{{\tt arXiv:2109.02653}}].

\bibitem{Altarelli:1977kt}
G.~Altarelli, G.~Parisi, and R.~Petronzio, {\it {Transverse Momentum in
  Drell-Yan Processes}},  {\em Phys. Lett. B} {\bf 76} (1978) 351--355.

\bibitem{Altarelli:1978pn}
G.~Altarelli, G.~Parisi, and R.~Petronzio, {\it {Transverse Momentum of Muon
  Pairs Produced in Hadronic Collisions}},  {\em Phys. Lett. B} {\bf 76} (1978)
  356--360.

\bibitem{Parisi:1978eg}
G.~Parisi, {\it {Super Inclusive Cross-Sections}},  {\em Phys. Lett. B} {\bf
  74} (1978) 65--67.

\bibitem{Parisi:1978jv}
G.~Parisi and N.~Sourlas, {\it {A Simple Parametrization of the $Q^2$
  Dependence of the Quark Distributions in {QCD}}},  {\em Nucl. Phys. B} {\bf
  151} (1979) 421--428.

\bibitem{Parisi:1978kw}
G.~Parisi and R.~Petronzio, {\it {Gluon fragmentation functions from quark
  jets}},  {\em Phys. Lett. B} {\bf 82} (1979) 260--262.

\bibitem{Parisi:1979se}
G.~Parisi and R.~Petronzio, {\it {Small Transverse Momentum Distributions in
  Hard Processes}},  {\em Nucl. Phys. B} {\bf 154} (1979) 427--440.

\bibitem{Parisi:1979xd}
G.~Parisi, {\it {Summing Large Perturbative Corrections in QCD}},  {\em Phys.
  Lett. B} {\bf 90} (1980) 295--296.

\bibitem{Collins:1984kg}
J.~C. Collins, D.~E. Soper, and G.~F. Sterman, {\it {Transverse Momentum
  Distribution in Drell-Yan Pair and W and Z Boson Production}},  {\em Nucl.
  Phys. B} {\bf 250} (1985) 199--224.

\bibitem{Angeles-Martinez:2015sea}
R.~Angeles-Martinez et~al., {\it {Transverse Momentum Dependent (TMD) parton
  distribution functions: status and prospects}},  {\em Acta Phys. Polon. B}
  {\bf 46} (2015), no.~12 2501--2534,
  [\href{http://xxx.lanl.gov/abs/1507.05267}{{\tt arXiv:1507.05267}}].

\bibitem{Altarelli:1979ub}
G.~Altarelli, R.~K. Ellis, and G.~Martinelli, {\it {Large Perturbative
  Corrections to the Drell-Yan Process in QCD}},  {\em Nucl. Phys. B} {\bf 157}
  (1979) 461--497.

\bibitem{Sterman:1986aj}
G.~F. Sterman, {\it {Summation of Large Corrections to Short Distance Hadronic
  Cross-Sections}},  {\em Nucl. Phys. B} {\bf 281} (1987) 310--364.

\bibitem{Catani:1989ne}
S.~Catani and L.~Trentadue, {\it {Resummation of the QCD Perturbative Series
  for Hard Processes}},  {\em Nucl. Phys. B} {\bf 327} (1989) 323--352.

\bibitem{Forte:2002ni}
S.~Forte and G.~Ridolfi, {\it {Renormalization group approach to soft gluon
  resummation}},  {\em Nucl. Phys. B} {\bf 650} (2003) 229--270,
  [\href{http://xxx.lanl.gov/abs/hep-ph/0209154}{{\tt hep-ph/0209154}}].

\bibitem{Contopanagos:1996nh}
H.~Contopanagos, E.~Laenen, and G.~F. Sterman, {\it {Sudakov factorization and
  resummation}},  {\em Nucl. Phys. B} {\bf 484} (1997) 303--330,
  [\href{http://xxx.lanl.gov/abs/hep-ph/9604313}{{\tt hep-ph/9604313}}].

\bibitem{Magnea:1990zb}
L.~Magnea and G.~F. Sterman, {\it {Analytic continuation of the Sudakov
  form-factor in QCD}},  {\em Phys. Rev. D} {\bf 42} (1990) 4222--4227.

\bibitem{Ahrens:2008qu}
V.~Ahrens, T.~Becher, M.~Neubert, and L.~L. Yang, {\it {Origin of the Large
  Perturbative Corrections to Higgs Production at Hadron Colliders}},  {\em
  Phys. Rev. D} {\bf 79} (2009) 033013,
  [\href{http://xxx.lanl.gov/abs/0808.3008}{{\tt arXiv:0808.3008}}].

\bibitem{Martin:2012da}
A.~D. Martin, A.~J. T.~M. Mathijssen, W.~J. Stirling, R.~S. Thorne, B.~J.~A.
  Watt, and G.~Watt, {\it {Extended Parameterisations for MSTW PDFs and their
  effect on Lepton Charge Asymmetry from W Decays}},  {\em Eur. Phys. J. C}
  {\bf 73} (2013), no.~2 2318, [\href{http://xxx.lanl.gov/abs/1211.1215}{{\tt
  arXiv:1211.1215}}].

\bibitem{Gao:2023ivm}
A.~Gao, H.~T. Li, I.~Moult, and H.~X. Zhu, {\it {The transverse energy-energy
  correlator at next-to-next-to-next-to-leading logarithm}},  {\em JHEP} {\bf
  09} (2024) 072, [\href{http://xxx.lanl.gov/abs/2312.16408}{{\tt
  arXiv:2312.16408}}].

\bibitem{Bonino:2024adk}
L.~Bonino, T.~Gehrmann, M.~Marcoli, R.~Sch\"urmann, and G.~Stagnitto, {\it
  {Antenna subtraction for processes with identified particles at hadron
  colliders}},  {\em JHEP} {\bf 08} (2024) 073,
  [\href{http://xxx.lanl.gov/abs/2406.09925}{{\tt arXiv:2406.09925}}].

\bibitem{Owens:1978qz}
J.~F. Owens, {\it {On the Q**2 Dependence of Parton Fragmentation Functions}},
  {\em Phys. Lett. B} {\bf 76} (1978) 85--88.

\bibitem{brezcont}
E.~Br\'ezin, {\it {Contribution to this volume}}, .

\bibitem{vulpcont}
A.~Vulpiani, {\it {Contribution to this volume}}, .

\bibitem{rychcont}
S.~Rychkov, {\it {Contribution to this volume}}, .

\end{thebibliography}\endgroup

\end{document}